\documentclass[graybox]{svmult}

\usepackage{mathptmx}       
\usepackage{helvet}         
\usepackage{courier}        
\usepackage{type1cm}        
\usepackage{makeidx}         
\usepackage{graphicx}        
\usepackage{multicol}        
\usepackage[bottom]{footmisc}
\usepackage{subfigure}
\usepackage{cite}

\makeindex             

\graphicspath{{./fig/}}

\begin{document}

\title*{A Universal Lifetime Distribution for Multi-Species Systems}
\author{Yohsuke Murase, Takashi Shimada, Nobuyasu Ito, and Per Arne Rikvold}
\institute{
  Yohsuke Murase and Nobuyasu Ito
  \at RIKEN Advanced Institute for Computational Science, 7-1-26, Minatojima-minami-machi, Chuo-ku, Kobe, Hyogo, 650-0047, Japan
\and
  Takashi Shimada and Nobuyasu Ito
  \at Department of Applied Physics, School of Engineering, The University of Tokyo, 7-3-1 Hongo, Bunkyo-ku, Tokyo 113-8656, Japan
\and
  Yohsuke Murase, Takashi Shimada, and Nobuyasu Ito
  \at CREST, Japan Science and Technology Agency 4-1-8 Honcho, Kawaguchi, Saitama, 332-0012, Japan
\and
  Per Arne Rikvold
  \at Department of Physics, Florida State University, Tallahassee, FL 32306-4350, USA
}
%
%
\maketitle

\abstract{
  Lifetime distributions of social entities, such as enterprises, products, and media contents, are one of the fundamental statistics characterizing the social dynamics.
  To investigate the lifetime distribution of mutually interacting systems, simple models having a rule for additions and deletions of entities are investigated.
  We found a quite universal lifetime distribution for various kinds of inter-entity interactions, and it is well fitted by a stretched-exponential function with an exponent close to 1/2.
  We propose a ``modified Red-Queen'' hypothesis to explain this distribution.
  We also review empirical studies on the lifetime distribution of social entities, and discussed the applicability of the model.
}

\section{Introduction}
\label{sec:intro}

Society is a system of diverse coexisting entities showing a high turnover of its membership \cite{sen2013sociophysics}.
Examples of such entities include enterprises, products, and media contents.
Lifetime distributions of these entities are one of the most fundamental properties of such systems.
Thus, understanding of these distributions will reveal the underlying social dynamics.
For example, the lifetime of products would be strongly related to the market trend,
and the lifetime of enterprises can be a crucial condition for the stability of economic activity and employment.
Although several models have been proposed to fit lifetime distributions,
most of these models do not explicitly take into account the interactions between entities.
Lifetime distributions of mutually interacting systems are not fully understood
even though these interactions often play a significant role in actual society as we see, for example, in cascading bankruptcies of enterprises.
In this short article, we investigate the lifetime distributions for ``ecosystem'' like systems, where diverse entities undergo competition for survival.

Several theoretical models have been proposed for lifetime distributions.
The simplest assumption is that an ``extinction'' (the elimination of an entity) occurs randomly with a constant rate, i.e., characterized by a Poisson process \cite{van1973new}.
The lifetime distribution of ``species'' (a social entity) is then a simple exponential function.
Although this assumption is mathematically simple, it is radical from a sociological point of view
because no evolutoinary advantage or aging effect is taken into account.
This hypothesis is known as the Red-Queen hypothesis or Van-Valen's law in evolutionary ecology.
On the other hand, if mortality rate is dependent on age, the lifetime distribution deviates from a simple exponential funciton \cite{Shimada:2003kx,lawless2011statistical}.
If a long-lived species has an evolutionary advantage, the probability that a species goes extinct will be a decreasing function of its age.
This assumption seems reasonable because a species which has been successful in surviving is expected to have some superior properties.
The decreasing mortality function yields a lifetime distribution with a heavier tail than the corresponding exponential function.
If a long-lived species has higher mortality than a younger species, perhaps due to its aging or degradation, the lifetime distribution will decay faster than the exponential one.
Another simple model is the return-time distribution of random walks \cite{van2007stochastic}.
Assuming that ``fitness'' of each species, which may be population or any other measure of the distance from extinction, follows a neutral random walk,
the lifetime distribution will be modeled by a return time distribution, which is known to be approximately $t^{-3/2}$.
If we assume a critical branching process instead of a random walk, we find a $t^{-2}$ power law.
A more general theory which combines a random walk and a branching process was also proposed \cite{pigolotti2005sld}.

All the above models consider a stochastic process of one species.
Lifetime distributions of mutually interacting species have been investigated mainly in the context of biological coevolution \cite{Drossel:2001lr,newman2003me}.
Among the simplest models of a mutually interacting system are the so-called self-organized criticality (SOC) models, which predict power laws \cite{bak93:_punct_equil_and_critic_in}.
In addition to these simplistic models, population dynamics models or individual based models are also proposed with the aim of bridging the ecological and evolutionary timescales.
These include the tangled-nature models \cite{PhysRevE.66.011904,CHRISTENSEN:2002yq,rikvold2003pea,0305-4470-38-43-005,Rikvold:2007lr,rikvold2007ibp,murase2010effects,murase2010random}, the web-world model \cite{Caldarelli:1998eu,drossel01:_influen_of_predat_prey_popul}, and others \cite{shimada-arob2002,tokita-tpb2003}.
All those models have population dynamics of each species (or birth-death process at an individual level) and rules for the emergence and extinctions of species.
Some of these show a power law lifetime distribution $t^{-2}$ \cite{rikvold2003pea,rikvold2007ibp,0305-4470-38-43-005,Rikvold:2007lr,murase2010effects,murase2010random}
while others show a curved line that lies somewhere between a power law and an exponential distribution: concave on a log-log plot and convex on a semi-log plot \cite{murase2010random,shimada-arob2002}.
Interestingly, these seem to be classified into a few universality classes regardless of the apparent diversity of the models \cite{murase2010random}.
In the models which add new species with randomly determined interaction coefficients, a skewed lifetime distribution is universally observed under various population dynamics equations.
This type of addition of new species is called ``migration'' because a new species is not correlated with the current species at all.
On the other hand, with the ``mutation'' model, where a new species appears as a result of a modification of a current species,
$t^{-2}$ power law is robustly observed.
Even though the models have quite different numbers of species, types of interactions, and network topologies, they share similar species-lifetime distributions implying the existence of universality.

In this article, we mainly focus on the skewed profile found for migration rules because it is the simplest model to add a new species.
We will show the origin of the skewed profile by introducing a simple graph model.
In the next section, the model definition and its typical results are given.
In Sect.~\ref{sec:mrq_hypo}, the origin of the skewed profile is explained using what we call the modified Red-Queen hypothesis.
Then, in Sect.~\ref{sec:disc} we review the empirical data observed in society and discuss their relation with the model.
The last section is devoted to conclusions.

\section{Dynamical Graph Model}
\label{sec:model}

In order to investigate the lifetime distributions of mutually interacting systems, we propose a simple dynamically evolving model which was originally introduced for biological community assembly \cite{murase2010simple}.
A system is represented by a weighted and directed network, which self-organizes by successive migrations and eliminations (extinctions) of nodes.
Each node $i$ has a state variable called ``fitness'' $f_i$, which is defined as the sum of the weights of incoming links, i.e., $f_i = \sum_j a_{ij}$, where $a_{ij}$ is the weight of a link from node $j$ to $i$.
Node $i$ can survive if its fitness is larger than or equal to zero, otherwise it is eliminated from the system.

\begin{figure}[ht!!]
  \begin{center}
    \subfigure{
      \includegraphics[height=.25\textwidth]{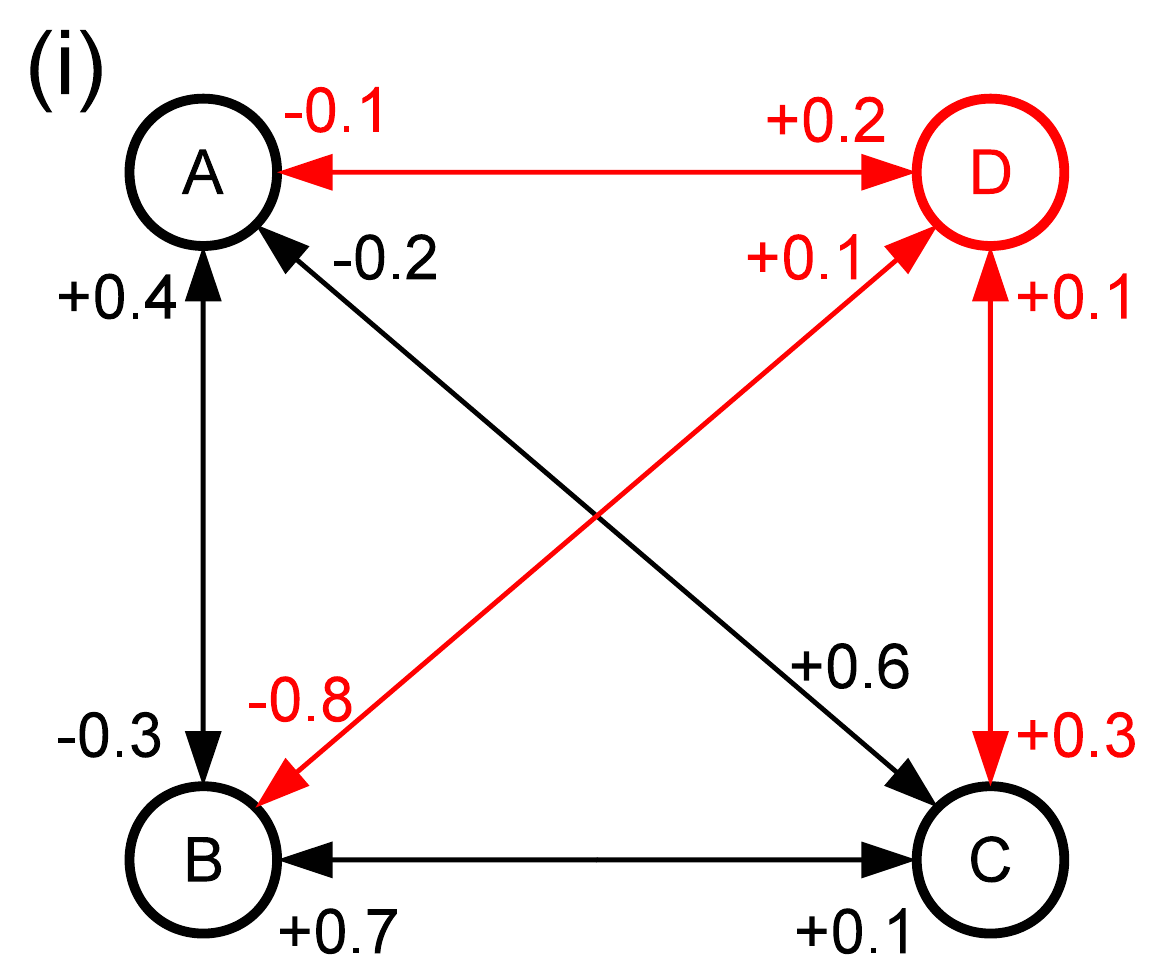}
    }
    \subfigure{
      \includegraphics[height=.25\textwidth]{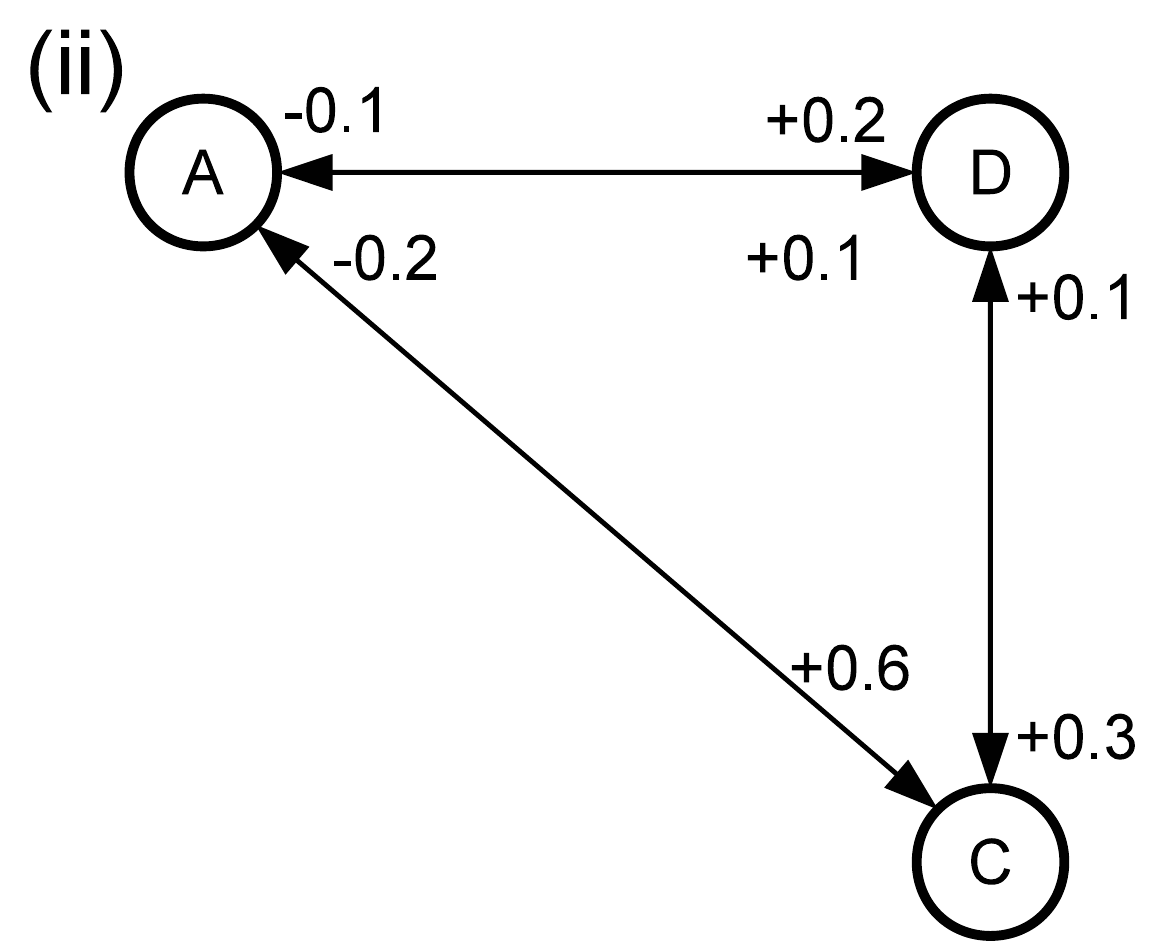}
    }
    \subfigure{
      \includegraphics[height=.25\textwidth]{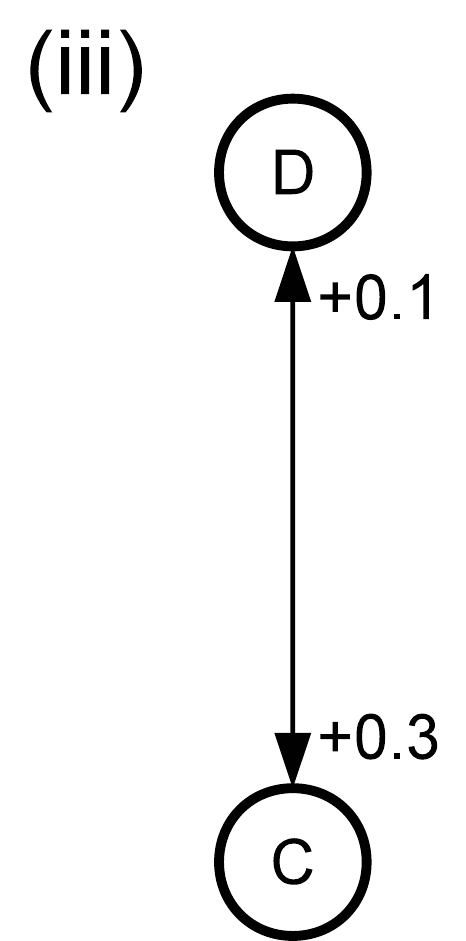}
    }
    \caption{
      (Color online) An example of the model dynamics.
      Nodes and arrows denote species and interactions, respectively.
      (i) Before the migration of species $D$, species $A$, $B$, and $C$ coexist.
      (ii)When species $D$ immigrates into the community, species $B$ goes extinct.
      (iii) Then, another resident species $A$ goes extinct due to the extinction of species $B$.
      After extinctions of species $A$ and $B$, all remaining species ($C$ and $D$) have positive fitness values.
      The figure is taken from \cite{murase2010simple}.
    }
    \label{fig:dg_model}
\end{center}
\end{figure}

At each time step, a new node is added to the system.
New links between existing nodes and the new node are randomly assigned with probability $c$, whose weights are randomly drawn from the Gaussian distribution with mean $0$ and variance $1$.
After a migration, the species with minimum fitness is identified and is eliminated from the system if the minimum fitness is negative.
Since the extinction of a node affects the fitness of other species, successive extinctions can happen.
This process is repeated until all the fitness values in the system become non-negative for each time step. (See Fig.~\ref{fig:dg_model}.)

\begin{figure}[ht!!]
\begin{center}
\includegraphics[width=.65\textwidth]{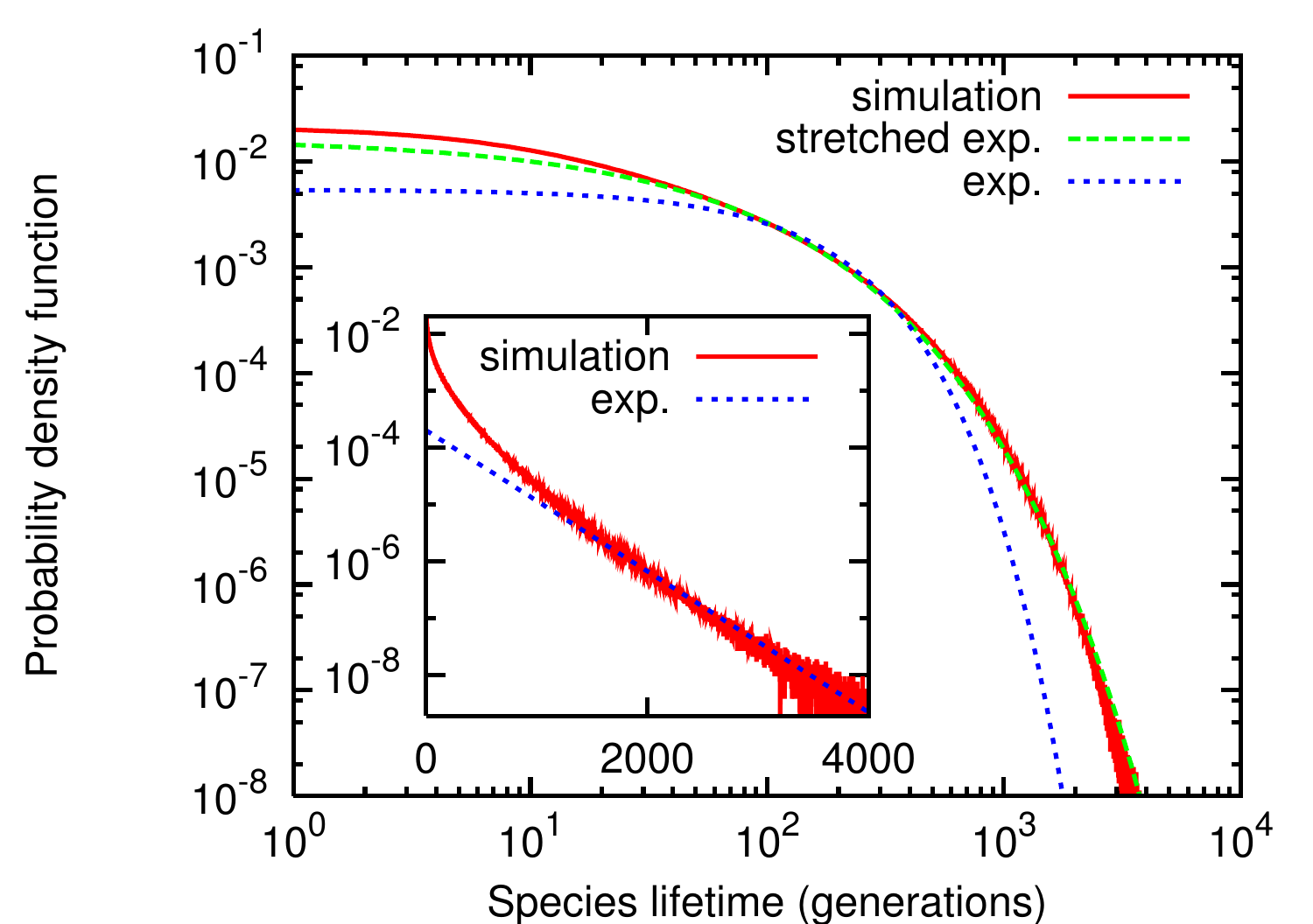}
\sidecaption
\caption{
(Color online) Species-lifetime distribution on log-log scales and linear-log scales (inset) for the dynamical graph model.
Fitting curves are also shown with a stretched exponential function and a simple exponential function.
In the inset, a fitting with a simple exponential function is also shown as a guide to the eye.
This figure is modified based on a figure in \cite{murase2010simple}.
}
\label{fig:lifetime}
\end{center}
\end{figure}

This simple model reproduces a characteristic skewed lifetime distribution found for population dynamics models with migration rules.
As shown in Fig.~\ref{fig:lifetime}, the distribution is neither a simple exponential nor a simple power law distribution.
It is well fitted by a stretched exponential function with an exponent close to 1/2.
Note that the number of species $N$ fluctuates in a finite range and the statistics are taken from a statistically stationary state.
Since this model shares a similar profile to ones for more complicated population dynamics models, this model is expected to capture the essential aspects of the skewed lifetime distribution.

\section{Modified Red-Queen Hypothesis}
\label{sec:mrq_hypo}

The lifetime distribution corresponding to a stretched exponential function with exponent $1/2$ is explained by what we call the modified Red-Queen hypothesis.
This hypothesis assumes that the mortality of each species is not dependent on its age but on the number of species in the system.
Let us assume that $N$ fluctuates in a finite range, and that the probabilities that $N$ increases or decreases are independent of $N$.
In other words, we assume that $N$ follows a random walk with a negative drift.
(Without a negative drift, we would get a divergence of $N$.)
These assumptions can be obtained by a mean-field analysis \cite{shimada2014universal,ohira2015mathematical}.
Based on these assumptions, the extinction probability of a species is proportional to $1/N$, meaning that the typical species lifetime $\tau$ is proportional to $N$.
Because a stable distribution of $N$ is an exponential function under these assumptions, a species-lifetime distribution is then expressed by a superposition of exponential functions of time scale $\tau$ with an exponential weight:

\begin{eqnarray}
  \label{eq:add-exponential}
  p(t) &=& \int_0^{\infty} \frac{\exp(-t/\tau)}{\tau} b\exp(-b\tau) d\tau \\
  &\approx& \sqrt{\pi}(bt)^{-1/4}\exp{(-2\sqrt{bt})} \qquad (t \gg 1),
\end{eqnarray}
where $b$ is a coefficient which depends on the probability that $N$ decreases.
This set of assumptions is called the modified Red-Queen hypothesis because the mortality is independent of the age of species.

\begin{figure}[ht!!]
\begin{center}
\includegraphics[width=.65\textwidth]{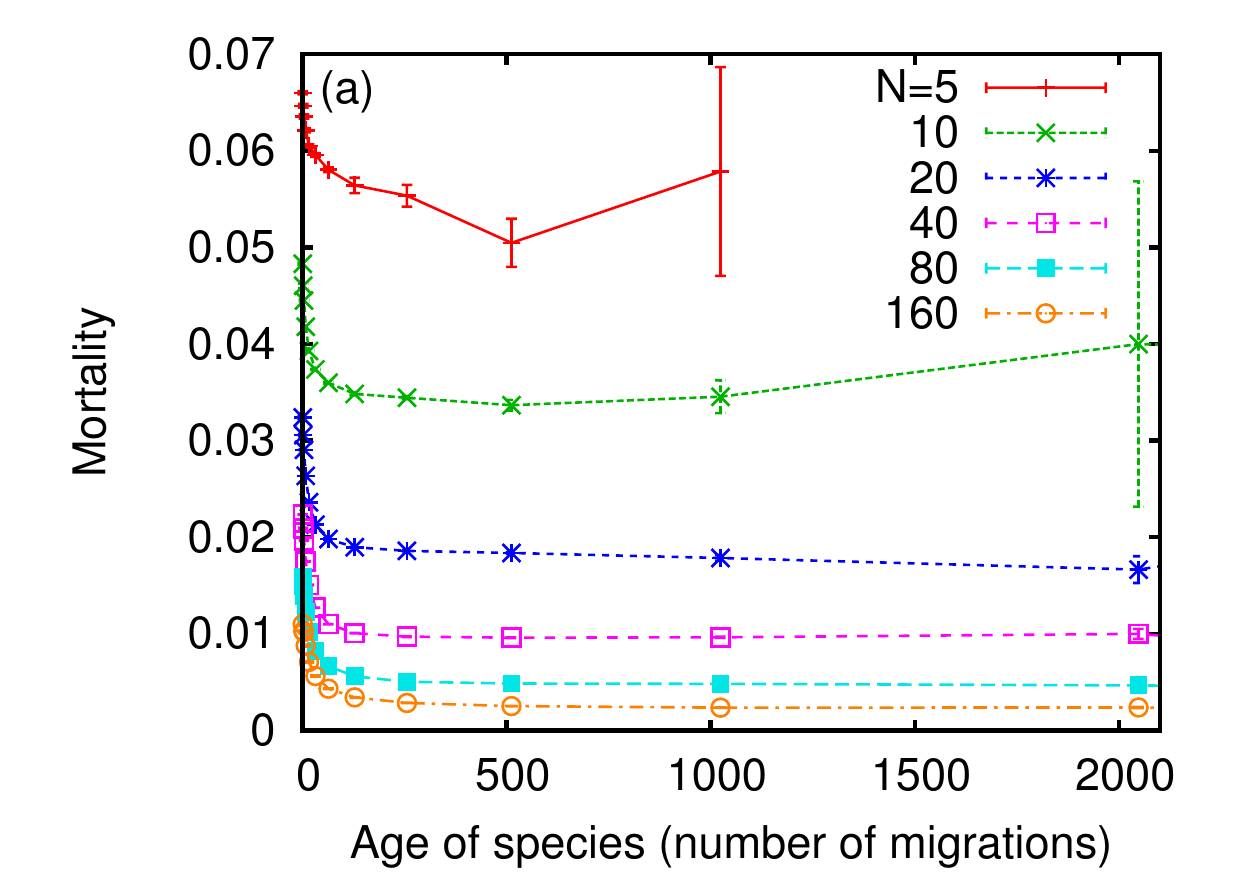}
\sidecaption
\caption{
(Color online)
Average probability that a species goes extinct (mortality) as a function of the age for communities of $N=5$, $10$, $20$, $40$, $80$, and $160$.
This figure is taken from \cite{murase2010simple}.
}
\label{fig:mortality}
\end{center}
\end{figure}

These assumptions are validated by calculating mortality against communities obtained by simulations.
We calculated the probability that each resident species goes extinct in the next time step by adding a test immigrant to the snapshot of the simulations.
The relationship between mortality and the age of a species is shown in Fig.~\ref{fig:mortality} for several numbers of species $N$.
For all $N$, mortalities show a sharp decrease at $t \approx 0$ and then converge to a constant value which is approximately proportional to $1/N$.
These results are in good agreement with the modified Red-Queen hypothesis.

\begin{figure}[ht!!]
\begin{center}
\subfigure{
  \includegraphics[height=.32\textwidth]{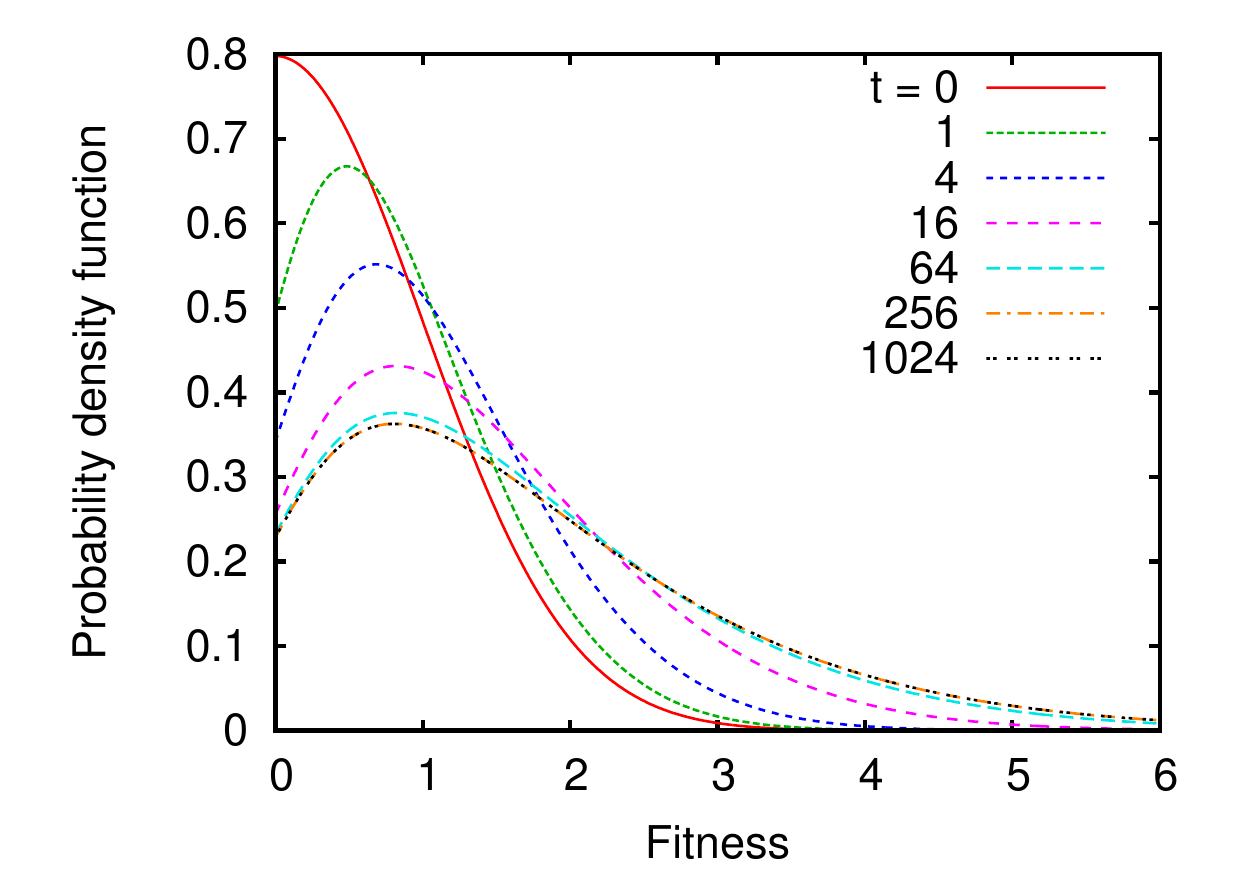}
}
\subfigure{
  \includegraphics[height=.32\textwidth]{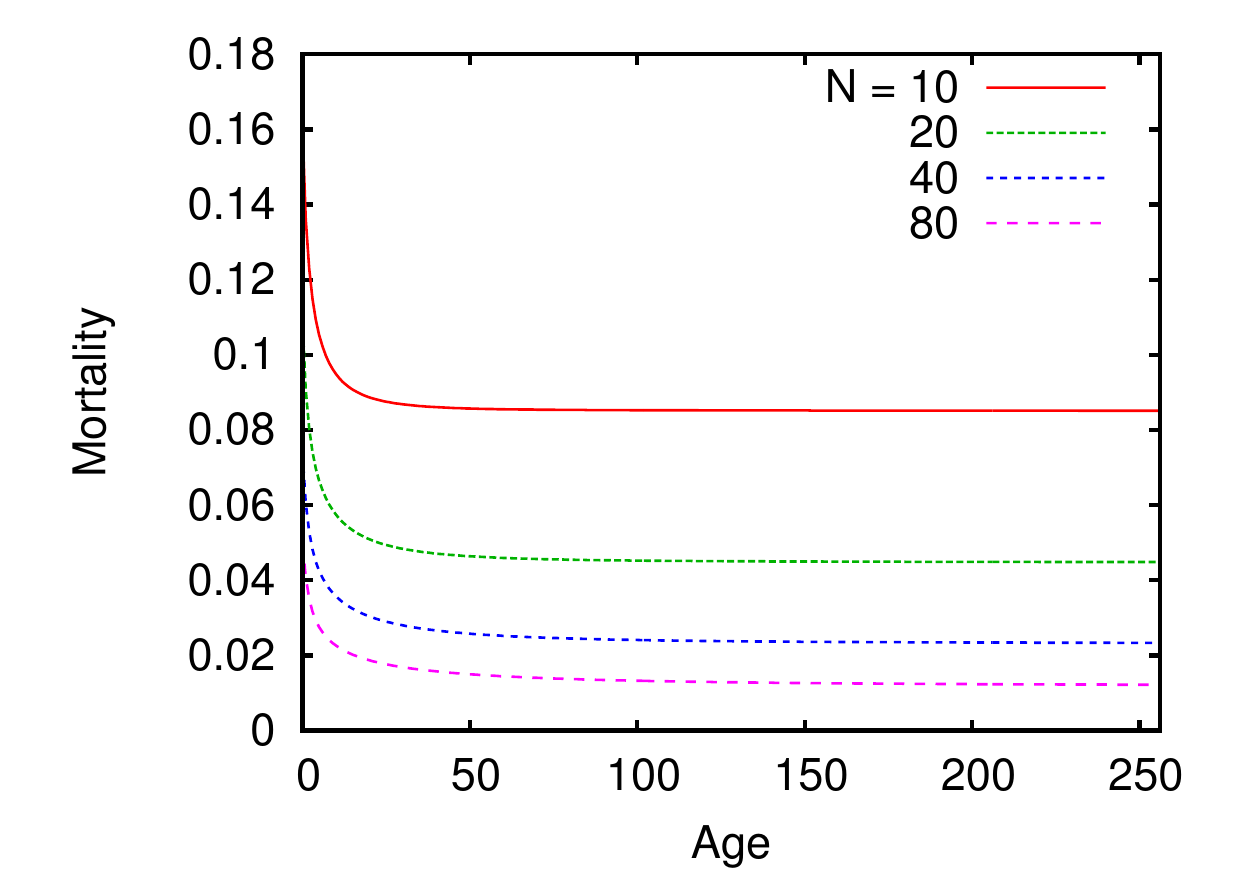}
}
\caption{
  (Color online)
  (left) Time evolution of the probability density function of the fitness, $P_{t}(f)$, under the fixed-$N$ approximation for $N=10$.
  (right) The probability of the species that goes extinct as a function of age. The probability is calculated under the fixed-$N$ approximation for several $N$.
}
\label{fig:fixed_n}
\end{center}
\end{figure}

The reason that the mortality does not depend on the age $t$ but on $N$ is explained by the time evolution of the fitness distribution.
Let us calculate the probability distribution of the fitness of a species at age $t$, $P_t(f)$,
under the assumption that $N$ is fixed, i.e., one immigrant comes to the system and one resident species goes extinct at each time step.
For simplicity, we discuss the case that $c = 1$.
The distribution of the fitness of a newcomer which succeeded in migrating, $P_0(f)$, is given by
\begin{eqnarray}
  \label{eq:pdf_f}
  P_0(f) = \left\{\begin{array}{ll}
        2 {\cal N}(0,N) & (f \geq 0) \\
        0 & (f < 0) \\
      \end{array}\right.,
\end{eqnarray}
where ${\cal N}(0, N)$ denotes the Gaussian distribution with mean $0$ and variance $N$, because the fitness is the sum of $N$ Gaussian random numbers.
At the next time step, the fitness changes due to the migration and the extinction.
While the interaction coefficient with an immigrant species is given by ${\cal N}(0,1)$,
the interaction coefficient with the resident species should have a positive mean because the sum of the incoming links is positive.
Therefore, we assume that the distribution of the interaction with the resident species is ${\cal N}(\mu_t/N, 1)$ where $\mu_t$ is the mean of $P_t(f)$.
In total, the distribution of total change in fitness for each time step is the convolution of ${\cal N}(0,1)$ and ${\cal N}(-\mu_t/N,1)$, i.e., ${\cal N}(-\mu_t/N,2)$.
Under this assumption, the time evolution of $P_t(f)$ is calculated as follows:
\begin{equation}
  \label{eq:pdf_f_time}
  P_{t+1}(f) = \left\{\begin{array}{ll}
      C_{t+1} \times \left[ P_{t} \ast {\cal N}(-\mu_t/N, 2) \right](f) & (f \geq 0) \\
      0 & (f < 0) \\
    \end{array}\right.,
\end{equation}
where $C_{t}$ is the normalization coefficients and the operator $\ast$ denotes the convolution.
The coefficient $C_t$ is determined so that the positive part of the convlolution function is normalized.
From this equation, mortality at each time step, $m(t)$ is also calculated as the ratio of the negative part of the convolution function.
Numerically calculated time evolution of $P_{t}(f)$ and $m(t)$ are shown in Fig.~\ref{fig:fixed_n}.
As $t$ increases, $P_{t}(f)$ and $m(t)$ approach a constant profile and a constant value, respectively.
The value to which $m(t)$ converges is inversely proportional to $N$.
All these are consistent with the simulation results and the modified Red-Queen hypothesis.

\begin{figure}[ht!!]
\begin{center}
  \includegraphics[width=0.6\textwidth]{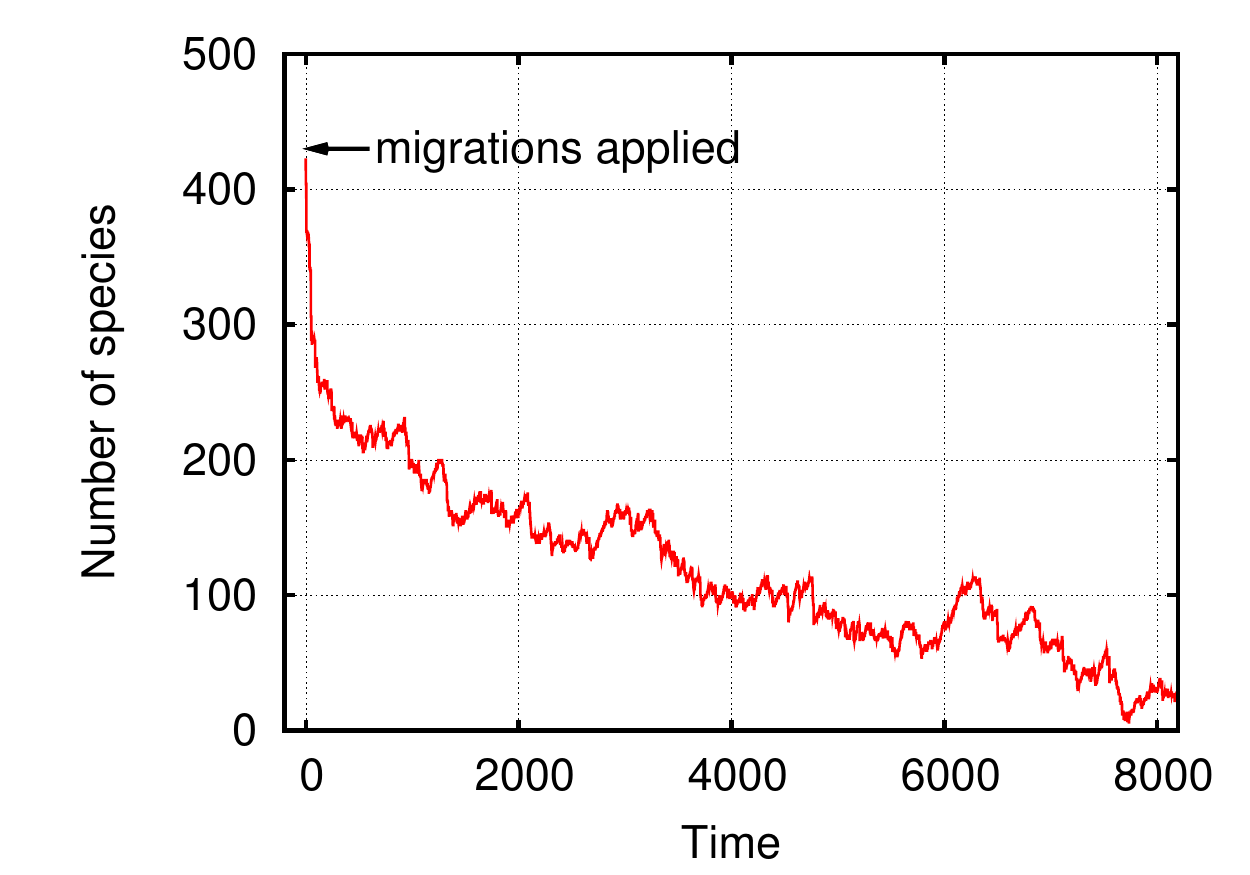}
  \sidecaption
  \caption{
    Time development of the number of species.
    A community with one thousand species whose interactions were randomly assigned was used as an initial state.
    The number of coexisting species was $423$ after eliminating the unfit species from the initial state.
    Then, migrations are applied. Due to the migrations, the number of species drops immediately.
    Connectance $c=0.2$ was used.
  }
  \label{fig:prepared_community}
\end{center}
\end{figure}

While the mortality converges to a constant value for large $t$, it shows a sudden drop at $t \sim 0$.
It means species that have just entered the system are more susceptible to extinctions than long-lived species.
This fact implies that the species compositions depend on the way they are constructed.
For example, let us prepare an initial state with one thousand species whose coupling constants are randomly assigned from a Gaussian distribution with mean zero.
After starting the dynamics of extinction, the community immediately loses approximately half of the species, while the rest of the species can coexist.
The community constructed in this way (here, we call it ``prepared'' community)
is qualitatively different from the communities constructed by repetitive migration-extinction processes (here, we call it ``trial-and-error'' community).
This difference is observed in the robustness against the migrations of new species.
Figure \ref{fig:prepared_community} shows the time evolution of $N$ for a prepared community which starts from $N=1000$.
Although $423$ species survived at $t=0$, most of these species went extinct as soon as migrations of new species were started.
Thus, prepared communities are fragile against migration of new species because the fitness distribution in prepared communities has a peak at around zero.
On the other hand, the trial-and-error communities are robust against migrations.
The number of species $N$ for the trial-and-error communities keeps fluctuating around a constant value under successive migrations.
This is an important implication that the model for the society should include a kind of trial-and-error construction process, oherwise we might miss an important point.

Finally, it should be emphasized that the modified Red-Queen hypothesis is quite different from the assumption of age-dependent mortality.
Actually, assuming the mortality function proportional to $t^{-1/2}$ can yield a Weibull distribution with exponent $1/2$, thus a similar lifetime distribution is obtained \cite{lawless2011statistical}.
However, these two assumptions have a crucial difference when predicting the extinction risk of a species.
The extinction risk can be estimated based on species age if an age-dependent mortality is valid, although this is not the case under the modified Red-Queen hyopthesis.
Furthermore, the modified Red-Queen hypothesis answers why the exponent is $1/2$ while the age-dependent mortality can assume an arbitrary exponent.

\section{Relation with Empirical Data}
\label{sec:disc}

In this section, we review several empirical data sets and discuss the applicability of our models.

The first example is product life cycles in a Japanese convenience store chain \cite{Mizuno:2009hb}.
It was found that the lifetime distributions of noodles, juice, and sandwiches show similar curves. 
Interestingly, these curves are fitted by the Weibull distribution with an exponent close to $1/2$, which is quite similar to the distribution found in our model.
Moreover, this skewed profile is found also in biological coevolution.
The data estimated from the fossil record show a skewed profile \cite{Shimada:2003kx}.
Although a $q$-exponential fit is proposed in \cite{Shimada:2003kx}, the stretched exponential function fits the data as well.
It indicates the universality shared by sociological and biological systems.

The second example is the lifetime distribution of Japanese firms.
The lifetime distribution of Japanese firms which went bankrupt in 1997 is well approximated by a simple exponential function \cite{Fujiwara:2004lr}, hence it seems consistent with the (non-modified) Red-Queen picture.
Thus, we can conclude that the age-dependent mortality picture is not valid at least for Japanese firms.
On the other hand, these data are not sufficient to reject the modified Red-Queen hypothesis even though the distribution is not skewed.
This is because the period of the measurement is only one year, which is much shorter than the typical lifetime.
A measurement on longer time scales will deepen our understanding of the bankruptcy dynamics of enterprises.
Another possible reason for the incosistency with our model is that the model assumes that the migration frequency of new species is independent of $N$.
If we assume an $N$-dependent migration rate, we will get a different lifetime distribution.
Data on the occurrence frequency of new enterprises would clarify this point.

Several lifetime distributions of media contents have also been investigated.
Media contents can be interacting with each other as most of these are competing for the attention of potential consumers.

The first data set of media contents is movie popularity \cite{pan2010statistical}.
The cumulative distribution of the persistence time of a movie fits a stretched exponential form with an exponent about $1.6$, indicating that it decays faster than exponentially.
This quick decay is explained by a few observed stylized facts and an assumptions that a movie is withdrawn when the gross income per theater gets below a threshold value.
The observed statistics shows a $1/t$ decay in gross income of a movie per theater, indicating that the mortality of a movie increases with age due to aging.
That is why we see a faster decay than a simple exponential function.

\begin{figure}[ht!!]
\begin{center}
  \includegraphics[width=0.62\textwidth]{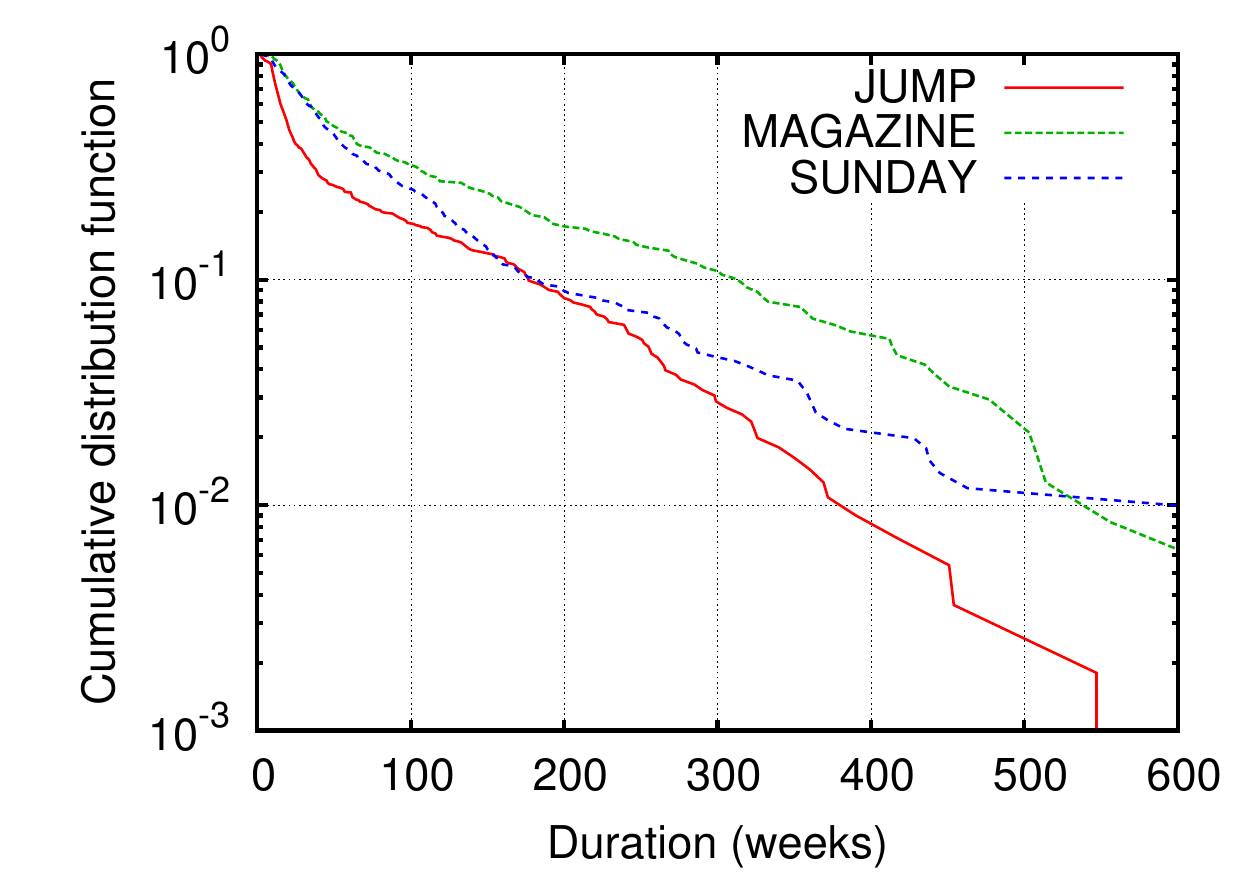}
  \sidecaption
  \caption{
    The cumulative distribution functions of lifetimes of the series
    that have run in Weekly Sh$\bar{\rm o}$nen Jump, Weekly Sh$\bar{\rm o}$nen Magazine, and Weekly Sh$\bar{\rm o}$nen Sunday.
    The lists were obtained from the Japanese pages of \cite{jump-list,sunday-list,magazine-list} in December 2009.
    Currently running series are not included in the statistics.
  }
  \label{fig:comic}
\end{center}
\end{figure}

Another example of media contents is comic series.
Figure \ref{fig:comic} shows the cumulative lifetime distributions of comic seies that have run in three major Japanese weekly comic magazines.
We defined the lifetime of each comic series as the duration between the first and the final issues, and collected data from a Wikipedia article which contains the list of the start and end dates for each series.
Since the termination of a series should strongly depend on its popularity, the series are competing with each other for a limited number of concurrent series.
As we see in the figure, all these distributions show approximate exponential decays.
Thus, the Red-Queen picture looks the most reasonable hypothesis for comic series.
This is clearly different from the movie duration.
We conjecture the reason of the difference is that the comic series has a new story every week while a movie is not renewed.
One of the possible reasons that the dynamical graph model is not applicable is that the number of concurrent series does not show fluctuations, which is a key for the modified Red-Queen hypothesis.

Another data set of persistence in social media can be found for Twitter \cite{asur2011trends}.
A recent study showed that the length of a topic sequence follows an approximate $t^{-2}$ power law.
This is the most heavy-tailed distribution among the above.
One of the characteristic points of Twitter trends is that a trend can be recurrent, i.e., the trends appear more than once.
The dynamical graph model clearly misses this point.
Therefore, the model is not appropriate for Twitter trends.

\section{Conclusion}
\label{sec:conclusion}

In this article, we investigated the lifetime distribution for an interacting systems.
A skewed lifetime distribution is robustly observed for a wide range of models, and it fits a stretched exponential function with exponent $1/2$.
We proposed the dynamical graph model and revealed that this profile is explained by the modified Red-Queen hypothesis.
This hypothesis is a novel idea and its meaning is completely different from the age-dependent mortality assumption even though both yield similar profiles.

We also reviewed empirical data sets and discussed the applicability of the model.
While some data sets are similar to our model, dissimilar data sets are also common.
It is clear that the dynamical graph model is one of the simplest models for mutually interacting systems, hence it is not applicable to all the data.
Based on this study, we expect further exploration both of theoretical models and empirical data.

A theoretically important question is to find other universality classes and identify key factors which change these classes.
For example, it is not clear what is the fundamental difference between the dynamical graph and the SOC models.
Effects of the network topology is another big open question.
The dynamical graph model is essentially described by an Erd\"{o}s-R\'{e}nyi random network, however real-world networks have internal structures.
Studies of scale-free networks or modular networks are expected.
We will get a much deeper insight into the dynamics of diverse and open systems, when these questions are answered.

\begin{acknowledgement}
  The systematic simulations in this study were assisted by OACIS \cite{Murase201473}.
  P.A.R. is supported in part by U.S. NSF Grant No.~DMR-1104829.
\end{acknowledgement}
%

\bibliographystyle{spphys}


\end{document}